\documentclass[twocolumn,prd,floatfix,preprintnumbers,nofootinbib,superscriptaddress, aps]{revtex4-1}
        
\usepackage{amssymb,amsmath,verbatim,mathtools,needspace,enumitem,etoolbox,graphicx,physics,microtype,afterpage,bm,soul}
\usepackage[dvipsnames]{xcolor}
\definecolor{linkcolor}{rgb}{0.0,0.3,0.5}
\usepackage[unicode, colorlinks=true, linkcolor=linkcolor, citecolor=linkcolor, filecolor=linkcolor,urlcolor=linkcolor, pdfusetitle]{hyperref}
\usepackage{orcidlink}
\usepackage[all]{hypcap}
\usepackage[T1]{fontenc}
\usepackage[utf8]{inputenc}
\usepackage{tabularx}
\usepackage{multirow}
\usepackage{cancel}
\renewcommand{\arraystretch}{1.4}

\graphicspath{{Plots/}}

\usepackage{lmodern}
\usepackage{ragged2e}
\allowdisplaybreaks
\usepackage{tikz}
\usepackage{color}
\usepackage{nicefrac}
\usepackage{framed}
\hypersetup{colorlinks, citecolor=darkblue, linkcolor=black, urlcolor=darkblue}
\definecolor{rossos}{cmyk}{0,1,1,0.55}
\definecolor{bluscuro}{rgb}{0.15, 0.2, .85}
\definecolor{bluchiaro}{cmyk}{1,.3,0.,0.1}
\definecolor{ForestGreen}{rgb}{0.13, 0.55, 0.13}
\definecolor{darkblue}{rgb}{0,0, 1.39}
\newcommand{\msun}{M_{\odot}}

\newcommand{\be}{\begin{equation}}

\newcommand{\ee}{\end{equation}}

\def\BH{\text{\tiny BH}}

\newcommand{\PBH}{\text{\tiny PBH}}
\newcommand{\NS}{\text{\tiny NS}}
\newcommand{\BNS}{\text{\tiny BNS}}
\newcommand{\GW}{\text{\tiny GW}}

\newcommand{\orb}{\text{\tiny orb}}

\def\lsim{\mathrel{\rlap{\lower4pt\hbox{\hskip0.5pt$\sim$}}
    \raise1pt\hbox{$<$}}}         
\def\gsim{\mathrel{\rlap{\lower4pt\hbox{\hskip0.5pt$\sim$}}
    \raise1pt\hbox{$>$}}}         

\newcommand{\jhu}{William H.\ Miller III Department of Physics and Astronomy, Johns Hopkins University, \\ 3400 North Charles Street, Baltimore, Maryland, 21218, USA}

\begin{document}

\title{Stochastic gravitational-wave background from subsolar neutron stars in collapsar disks}

\author{Valerio De Luca\orcidlink{0000-0002-1444-5372}}
\email{vdeluca2@jh.edu}
\affiliation{\jhu}

\author{Loris Del Grosso\orcidlink{0000-0002-6722-4629}}
\email{ldelgro1@jh.edu}
\affiliation{\jhu}

\author{Emanuele Berti\orcidlink{0000-0003-0751-5130}}
\email{berti@jhu.edu}
\affiliation{\jhu}

\begin{abstract}
\medskip
\noindent
The fragmentation of the neutrino-cooled accretion disks that surround the black holes formed in collapsars has recently been proposed as an astrophysical channel for the production of neutron stars with masses well below the standard core-collapse floor. These disk-born objects merge either with one another or with the central black hole, and the superposition of all such mergers across cosmic history sources a stochastic gravitational-wave background. In this work we compute the energy density of this background for both merger channels, adopting a formation history tied to the observed rate of long gamma-ray bursts. We find that the merger of a subsolar neutron star with the central black hole is the loudest channel, already constrained by current data and within reach of next-generation observatories, while the binary neutron-star channel is at most two orders of magnitude fainter, with its widest binaries remaining accessible to next-generation detectors. Confronting these predictions with the fourth observing run of the LIGO--Virgo--KAGRA Collaboration, we derive two upper bounds on the fraction of collapsars whose disks produce such mergers: a robust one from the non-detection of an isotropic background,  and a tighter one that could be achieved by a dedicated search for resolvable subsolar events in current data. Gravitational waves from non-axisymmetric deformations of the collapsar disk itself would provide an additional independent signature for the collapsar scenario, enabling cross-correlation studies with future deci-Hz detectors. Overall, the background computed here provides a population-level probe of disk-born subsolar neutron stars, which next-generation detectors could bring firmly into view.
\end{abstract}

\preprint{ET-0763A-26}
\maketitle

\clearpage

\section{Introduction}
\label{sec:intro}
\noindent
The detection of a compact object with a mass below the Chandrasekhar limit
$\sim 1\,\msun$ would be one of the most consequential outcomes of the current
gravitational-wave (GW) observational program~\cite{LIGOScientific:2018mvr, LIGOScientific:2020ibl, LIGOScientific:2021usb, KAGRA:2021vkt, LIGOScientific:2025hdt, LIGOScientific:2025slb, LIGOScientific:2026wfs}. Standard stellar evolution produces neither
remnant black holes (BHs) nor ordinary core-collapse neutron stars (NSs)
in this range~\cite{Chandrasekhar:1931ih,Shapiro:1983du,Burrows:2020qrp,Lattimer:2021emm, Karim:2026bva},
so a confident subsolar detection would point either to new physics or to an
unconventional formation pathway~\cite{Shandera:2018xkn, Singh:2020wiq, Essick:2024olf, Krnjaic:2026sgl, Gao:2026zwc}. The LIGO--Virgo--KAGRA (LVK) Collaboration has
carried out dedicated searches for subsolar coalescences across its observing
runs, placing upper limits on their rate and reporting no confident
detection~\cite{Abbott:2018oah,Nitz:2020bdb, Nitz:2021mzz, LIGOScientific:2021job,Phukon:2021cus, Nitz:2022ltl,LVK:2022ydq,Kacanja:2024hme, Soni:2024cdb,Kacanja:2026byy,LIGOScientific:2026wxz}. During the fourth observing run (O4) it has issued public low-latency alerts for two subsolar event candidates, S250818k and S251112cm~\cite{2025GCN.41437....1L, ligo_scientific_collaboration_ligovirgokagra_2025}, which triggered extensive electromagnetic follow-up~\cite{Kasliwal:2025keb, Franz:2025uan, Gillanders:2025fwf, Yang:2025dtj, Ackley:2026zlf, Vieira:2026eof, ODwyer:2026caq, Hall:2026gov, Tejera:2026dpr, Liu:2026hmk}. Next-generation detectors, such as the Einstein
Telescope (ET)~\cite{Hild:2010id, Punturo:2010zz,Branchesi:2023mws,ET:2025xjr} and Cosmic
Explorer (CE)~\cite{Reitze:2019iox,Evans:2021gyd, Evans:2023euw}, are expected to observe these sources out to cosmological distances.
 
The subsolar range has long been regarded as a signpost of primordial BHs
(PBHs), i.e., BHs formed in the radiation-dominated era from the collapse of large
density perturbations~\cite{Zeldovich:1967lct, Hawking:1971ei, Carr:1974nx, Carr:1975qj}, which produce subsolar binaries through early-Universe
pairing or dynamical capture (see Refs.~\cite{Sasaki:2018dmp,Carr:2020gox,Green:2020jor,LISACosmologyWorkingGroup:2023njw,Byrnes:2025tji}
for reviews). In the $(0.01-1)\,\msun$ window, PBHs can make up at most $\sim1\%$
of the dark matter, being tightly bounded by microlensing surveys and GW
searches~\cite{Carr:2020gox,Green:2020jor,Carr:2026hot}, yet they may still leave a
detectable imprint~\cite{Prunier:2023uoo,Markin:2023fxx,Yuan:2024yyo,Pujolas:2021yaw,DeLuca:2021hde,Franciolini:2023opt,Miller:2024rca,Magaraggia:2026jhk,Christensen:2026qfx}.
Because their waveforms are those of point masses in vacuum, disentangling a
genuinely astrophysical origin from a primordial one is central to the
interpretation of any such trigger, and the two populations must ultimately be
told apart by features beyond the
masses~\cite{Cardoso:2017cfl,Cardoso:2019rvt,Cardoso:2019upw,Barsanti:2021ydd,Franciolini:2021tla,Franciolini:2021xbq,Cole:2022fir,Coogan:2021uqv,Crescimbeni:2024cwh,Crescimbeni:2024qrq,DeLuca:2024uju,Russo:2025ivk,DeLuca:2025bph,Begnoni:2025aqc,Corman:2026lbt}.
 
A physically motivated astrophysical alternative to a primordial origin has
recently emerged in the form of NSs born in the outer regions of collapsar
disks~\cite{Piro:2006ja, Wu:2023qeh, Metzger:2024ujc,Lerner:2025dkd,Chen:2025uwd}. In the collapsar
picture, the core collapse of a rapidly rotating massive star produces a
stellar-mass BH surrounded by a dense, neutrino-cooled accretion disk that can power
long gamma-ray bursts (LGRBs)~\cite{MacFadyen:1998vz,Woosley:2006fn}. If the disk
carries enough mass at large radii, its outer regions become gravitationally
unstable and fragment into neutron-rich clumps that cool and contract into
subsolar-mass NSs. These objects then evolve dynamically within the disk,
producing two sequential classes of GW sources: mergers of two disk-born NSs (a
binary NS, or BNS, orbiting the central BH as a hierarchical triple), and
the inspiral of a surviving NS remnant into the central BH itself (a NS-BH
system)~\cite{Baibhav:2026zuy}. The individual-event signatures of these mergers
have been characterized in detail in a companion paper~\cite{Baibhav:2026iny}.

An observable complementary to, and largely independent of, the individual mergers is the
stochastic GW background (SGWB): the incoherent superposition of all sources too
weak or too distant to be resolved individually, which forms a persistent signal
encoding the integrated emission history of a population~\cite{Christensen:2018iqi,Regimbau:2011rp,Renzini:2022alw}.
Its energy density spectrum $\Omega_\GW(f)$ carries information both on the redshift-integrated
merger rate and on the spectral shape of the underlying sources, and thus provides
population-level insights that resolved events alone cannot. In the band of
ground-based detectors, the dominant astrophysical contribution is expected to
come from unresolved super-solar compact binary coalescences~\cite{LIGOScientific:2016fbo,Regimbau:2011rp},
whose inspiral-dominated emission produces the characteristic power law
$\Omega_\GW\propto f^{2/3}$; subdominant contributions are anticipated from
core-collapse supernovae and from rotating NSs and magnetars~\cite{Christensen:2018iqi}.
The LVK Collaboration has searched for an isotropic background in each observing
run, setting upper limits that are now within a factor of a
few of the predicted compact-binary background~\cite{KAGRA:2021kbb,Virgo:2025aai}.
A new class of subsolar sources would add its own contribution to this
background, with an amplitude fixed by its rate and a spectral shape set by its
characteristic masses, making the SGWB a natural probe of the properties of this
channel.
 
Here we compute the stochastic imprint of the disk-born subsolar
population. Summed over all redshifts, the two merger channels source a SGWB whose
amplitude is set by the collapsar formation history and by the fraction of
collapsars whose disks fragment; the NS-BH merger is the loudest, with the BNS
channel up to about two orders of magnitude below but still within
next-generation reach. Confronting the prediction with the most recent LVK
bound on the isotropic SGWB~\cite{Virgo:2025aai} and with the undetected subsolar events, we derive two complementary
upper bounds on the collapsar fraction --- a robust one from the non-detection
of the background and a tighter one that could be achieved by a dedicated search for resolvable subsolar events. We also compare the predicted background with that from PBHs. We adopt
geometrical units ($G=c=1$) and the notation of the companion
paper~\cite{Baibhav:2026iny}.

\section{Subsolar neutron stars in collapsar disks}
\label{sec:model}
\noindent
We collect here the elements of the collapsar-disk scenario needed to build the
background, following Ref.~\cite{Baibhav:2026iny}, to which we refer for a
complete treatment.
 
\subsection{The collapsar disk model}
\label{sec:disk}
\noindent
The core collapse of a rapidly rotating massive star leaves behind a stellar-mass BH,
$M_\BH\lesssim\mathcal{O}(30)\,\msun$, surrounded by a hot, neutrino-cooled
accretion disk sustained by an accretion rate of $\dot{M}_\BH\sim(0.1-1)\,\msun\,{\rm s}^{-1}$~\cite{MacFadyen:1998vz,Woosley:2006fn,Gottlieb:2023cgm, Metzger:2024ujc}.
Modeling the outer disk as a steady-state $\alpha$-disk~\cite{Shakura:1972te}, 
with viscosity $\alpha\simeq0.03$~\cite{Siegel:2018zxq} and vertical aspect
ratio $\mathbb{H}=h/r$, the surface and midplane densities follow the profiles
\begin{equation}
    \Sigma(r)=\frac{\dot{M}_\BH}{3\pi\alpha\mathbb{H}^2}\frac{1}{\sqrt{M_\BH r}}\,,
    \qquad
    \rho(r)=\frac{\Sigma}{2h}\propto r^{-3/2}\,.
    \label{eq:profiles}
\end{equation}
The outer disk becomes gravitationally unstable where its Toomre parameter
$Q=c_s\kappa/\pi\Sigma$, where $c_s$ is the midplane sound speed, and $\kappa\simeq\Omega\equiv\sqrt{M_\BH/r^3}$  is the epicyclic
(orbital) frequency of the Keplerian disk, drops to
$Q_0\simeq\mathcal{O}(1)$~\cite{Toomre:1964zx,Lodato:2004wf}, provided cooling is
fast enough that thermal pressure cannot stabilize the collapse
($\beta\equiv\tau_\text{\tiny cool}\Omega\lesssim\beta_c\sim10$~\cite{2021MNRAS.503.4192B}) --- a condition
amply satisfied by the copious neutrino emission of the
disk~\cite{Gammie:2001bw,Beloborodov:2002af,Lerner:2025dkd}. Fragmentation sets in near
\begin{equation}
    R_{*}\simeq \frac{62}{Q_0^{2/3}}\,R_g
    \left(\frac{\mathbb{H}}{0.3}\right)^{2}
    \left(\frac{\dot{M}_\BH}{\msun/{\rm s}}\right)^{-2/3}\,,
    \label{eq:Rstar}
\end{equation}
with $R_g\equiv GM_\BH$, and the unstable region fragments into $N_\NS\simeq2/\mathbb{H}\sim\mathcal{O}(10)$ clumps~\cite{Chen:2025uwd}, of characteristic mass
\begin{equation}
    m_\NS\simeq \frac{\mathbb{H}^3}{Q_0}\,M_\BH
    \simeq 0.1\,\msun\left(\frac{\mathbb{H}}{0.3}\right)^{3}
    \left(\frac{M_\BH}{5\,\msun}\right)\,.
    \label{eq:clumpmass}
\end{equation}
Because the weak interactions that cool the gas also neutronize it, driving the
electron fraction to $Y_e\lesssim0.1$, the local Chandrasekhar mass
$M_\text{\tiny Ch}\propto Y_e^2$ is well below $\msun$, and the clumps cool and
contract into stable subsolar-mass NSs~\cite{Lattimer:2004pg,Chen:2025uwd}. Their
structure is set by the cold equation of state (EoS) prevailing after collapse; throughout we adopt as
benchmark the non-relativistic neutron polytrope, defined
across the entire subsolar range~\cite{Metzger:2024ujc}, for which
\begin{equation}
    R_\NS\simeq 33\,{\rm km}\,(1-Y_e)^{-1/3}
    \left(\frac{m_\NS}{0.1\,\msun}\right)^{-1/3}\,.
    \label{eq:Rpoly}
\end{equation}
More realistic EoSs, such as ordinary nuclear-matter and quark-matter
models~\cite{Chodos:1974je, Johnson:1975zp, Prakash:1995uw, Akmal:1998cf, Douchin:2001sv}, are examined in the companion paper~\cite{Baibhav:2026iny}, to which we
refer for a detailed discussion. Fragmentation consumes most of the outer disk,
leaving a residual density $\rho_\text{\tiny env}=f_\text{\tiny gas}\,\rho(R_*)$
with a strongly uncertain suppression $f_\text{\tiny gas}\approx10^{-6}-10^{-2}$~\cite{Chen:2025uwd},
which we treat as a free parameter. The resulting population --- subsolar NSs of
mass $m_\NS\sim (0.01-1)\,\msun$ orbiting a central BH of mass
$M_\BH\sim(3-30)\,\msun$ --- feeds the two merger channels below, each of which
sources the background through its own characteristic emission.
 
\subsection{The BNS channel}
\label{sec:bns}
\noindent
Two disk-born NSs can pair into an inner binary that orbits the central BH,
making the system a hierarchical triple. The full evolution is an $N$-body
problem involving hierarchical mergers, recoil kicks, and gas
torques~\cite{Tagawa:2019osr, Varma:2020nbm, Metzger:2024ujc,Lerner:2025dkd}; rather than modeling it in detail,
we isolate the representative merger of two equal-mass NSs of mass $m_\NS$ and
initial separation $a_{\BNS,i}$. Such a binary may form either by fission
of an over-spinning clump, which deposits excess angular momentum into a close,
mildly eccentric pair~\cite{Alexander:2008wv,Wu:2026hth}, or by capture of two
independently formed NSs, favored in the gas-rich environment where gas dynamical
friction~\cite{Chandrasekhar:1943ys, Ostriker:1998fa, Kim:2007zb, Kim:2008ab} enhances the cross section~\cite{Chen:2025uwd}. We treat $m_\NS$ and
$a_{\BNS,i}$ as free parameters, the latter bounded from above by the Hill radius of
the pair relative to the central BH~\cite{Baibhav:2026iny}, located at an initial distance $a_{\BNS-\BH,i}$,
\begin{equation}
    r_\text{\tiny Hill}=a_{\BNS-\BH,i}\left(\frac{2m_\NS}{3M_\BH}\right)^{1/3}
    \simeq R_*\left(\frac{2m_\NS}{3M_\BH}\right)^{1/3}\,,
    \label{eq:rHill}
\end{equation}
and parametrized through the dimensionless Hill fraction
$f_\text{\tiny H}\equiv a_{\BNS,i}/r_\text{\tiny Hill}\in(0,1)$~\cite{Baibhav:2026iny}.
 
The subsequent evolution is governed by the competition of three timescales: the
dynamical friction hardening time in the residual gas~\cite{Baibhav:2026iny}
\begin{align}
    \tau_\text{\tiny dyn}
    &\simeq 0.3\,{\rm s}
    \left(\frac{M_\BH}{5\,\msun}\right)
    \left(\frac{\mathbb{H}}{0.3}\right)^{3}
    \left(\frac{f_\text{\tiny H}}{0.5}\right)^{-3/2} \nonumber \\
    &\times\left(\frac{\dot{M}_\BH}{\msun\,{\rm s}^{-1}}\right)^{-1}
    \left(\frac{f_\text{\tiny gas}}{10^{-2}}\right)^{-1},
    \label{eq:taudyn}
\end{align}
the Peters GW inspiral time~\cite{Peters:1963ux,Peters:1964zz} 
\begin{align}
    \tau_\GW^\BNS=\frac{5}{512}\frac{a_\BNS^4}{m_\NS^3}
    &\simeq 1\,{\rm day}\,
    \left(\frac{f_\text{\tiny H}}{0.5}\right)^{4}
    \left(\frac{m_\NS}{0.1\,\msun}\right)^{-5/3}
     \nonumber \\
    &\times
    \left(\frac{\rho(R_*)}{10^8 {\rm g/cm^3}}\right)^{-4/3}\,,
    \label{eq:tauGW}
\end{align}
for which GW emission removes an order-unity fraction of the orbital energy,
and the outer orbital period $\tau_\orb=2\pi/\Omega$ about the central BH. Their
ordering partitions the parameter space into regions, sketched in
Fig.~5 of Ref.~\cite{Baibhav:2026iny}: where the residual gas is dense,
$\tau_\text{\tiny dyn}<\tau_\GW^\BNS$, dynamical-friction drag extracts orbital energy faster than
GW emission and drives the pair to contact --- a gas-driven regime whose extent
grows with $f_\text{\tiny gas}$ [Eq.~\eqref{eq:taudyn}]. Where the gas is
tenuous, $\tau_\text{\tiny dyn}>\tau_\GW^\BNS$, the inspiral is GW-driven, and
the further comparison with $\tau_\orb$ splits this region in two. If the binary
merges within a fraction of an outer orbit, $\tau_\GW^\BNS\ll\tau_\orb$, the
triple geometry is frozen and the emission is that of a clean, isolated inspiral
proceeding to contact; we call this the ``GW regime''. If instead
$\tau_\GW^\BNS\gtrsim\tau_\orb$, the pair completes many revolutions about the
central BH before merging, and the outer orbit carries its imprint on the
waveform; we call this the ``modulation regime''. Since $f_\text{\tiny gas}$
is the most uncertain parameter and $\tau_\text{\tiny dyn}$ scales inversely with
it, these boundaries shift by orders of magnitude across the plausible range, so
we retain the two GW-driven regimes as the limiting cases that bracket the BNS
emission. For each $(M_\BH,\,m_\NS)$ we place the binary at the widest separation compatible with each regime, which maximizes its in-band emission~\cite{Baibhav:2026iny}: the edge of the GW-driven region $\tau_\text{\tiny dyn}=\tau_\GW^\BNS$ in the modulation regime, and the condition $\tau_\GW^\BNS=\tau_\orb$ in the GW regime.
 
Several effects then operate on the GW-driven inspiral, with different relevance in the two regimes. In the modulation regime the outer orbit imprints kinematic
and propagation modulations~\cite{Meiron:2016ipr,Chen:2018axp,Chen:2020iky,Toubiana:2020drf,Cardoso:2021vjq,Sberna:2022qbn,Yin:2024nyz,Santos:2025ass,Santos:2026lzq,Cardoso:2026ugm} that leave the emitted energy
nearly unchanged, so their effect on the background is indirect and twofold~\cite{Baibhav:2026iny}: (i) the wide
outer orbits of this regime lower the entry frequency, given by
\begin{equation}
    f^\BNS_i=\frac{1}{\pi}\sqrt{\frac{2m_\NS}{a_{\BNS,i}^{3}}}
    \simeq 3.6 \, {\rm Hz}\,f_\text{\tiny H}^{-3/2}
    \left(\frac{\rho(R_*)}{10^{8}\,{\rm g\,cm^{-3}}}\right)^{1/2}\,,
    \label{eq:fi}
\end{equation}
where the last equality assumes $a_{\BNS-\BH,i}=R_*$ and uses
Eqs.~\eqref{eq:Rstar} and~\eqref{eq:rHill}; and (ii) they
expose the binary to tidal unbinding by the central BH, which can truncate the
inspiral before contact and shifts the emission to lower frequencies. The pair
is stripped once the shrinking inner-binary Hill radius falls below the inner
separation, which maps to the unbinding frequency~\cite{Baibhav:2026iny}
\begin{equation}
    f_\text{\tiny unb} = f^\BNS_i
    \left\{1-\frac{3^{4/3}\left(2q\right)^{2/3}\left(1-f_\text{\tiny H}^{4}\right)}
    {f_\text{\tiny H}^{4}\left[4 - 3^{4/3}\left(2q\right)^{2/3} + 8q\right]}\right\}^{-3/8}\,,
    \label{eq:funb}
\end{equation}
as a function of the mass ratio $q\equiv m_\NS/M_\BH \ll 1$ and $f_\text{\tiny H}$. Equation~\eqref{eq:funb} follows the joint radiation-reaction decay of the inner and outer orbits, so that the outer orbit keeps shrinking rather than freezing~\cite{Baibhav:2026iny}. We evaluate  the background at the entry frequency $f^\BNS_i$ and at the Hill fraction marking the onset of the modulation regime.
At fixed $q$, Eq.~\eqref{eq:funb} interpolates between two limits. For $f_\text{\tiny H}\to1$ the pair is born at the edge of its Hill sphere and is stripped almost immediately, $f_\text{\tiny unb}\simeq f^\BNS_i$. As $f_\text{\tiny H}$ decreases, the inner binary shrinks faster than its Hill radius and $f_\text{\tiny unb}$ grows, diverging at a critical Hill fraction $\simeq (2q)^{1/6}$; tighter binaries are never unbound and radiate up to contact.

In both regimes the internal structure of the stars matters~\cite{Baibhav:2026iny}. Their enormous tidal
deformabilities --- of order $10^{9}$ for a $0.1\,\msun$ NS~\cite{Hinderer:2007mb, Flanagan:2007ix}, far larger than for
an ordinary NS --- accelerate the late inspiral, and in the gas-driven regime
dynamical friction adds a negative post-Newtonian term that further
speeds up the frequency evolution~\cite{Barausse:2014tra,Cardoso:2019rou, CanevaSantoro:2023aol}. For the background  these enter both through the in-band cycle count and the termination frequency~$f_\text{\tiny TD}=\min(f_\text{\tiny Roche},f_\text{\tiny unb})$, i.e., the lower of the frequency at which the pair reaches contact, $f_\text{\tiny Roche}$, and that at which the central BH
tidally unbinds it, $f_\text{\tiny unb}$. For an equal-mass BNS the Roche (contact) frequency reads
\begin{equation}
    f_\text{\tiny Roche}^\BNS=\frac{1}{\pi}\sqrt{\frac{m_\NS}{4R_\NS^3}}
    \simeq 90\,{\rm Hz}\left(\frac{m_\NS}{0.1\,\msun}\right)\,.
    \label{eq:fRocheBNS}
\end{equation}
This frequency sets the termination condition in the GW regime, where the binary radiates all the way to contact, while in the modulation regime it may be unbound earlier if $f_\text{\tiny unb}<f_\text{\tiny Roche}^\BNS$. Since $f_\text{\tiny unb}$ decreases with $f_\text{\tiny H}$, the two regimes populate distinct, partly overlapping frequency ranges.

\subsection{The NS-BH channel}
\label{sec:nsbh}
\noindent
The second channel is the inward migration and eventual merger of a surviving NS
--- a single clump, or the remnant of a hierarchical BNS merger
chain~\cite{Tagawa:2019osr, Metzger:2024ujc} --- with the central BH. The NS enters the band at the GW frequency of its birth orbit, which, using the marginal-stability condition $\Omega^2(R_*)=2\pi Q_0\,\rho(R_*)$, reads
\begin{equation}
    f_i^\text{\tiny NS-BH}=\frac{1}{\pi}\sqrt{\frac{M_\BH(1+q)}{R_*^{3}}}
    \simeq 2\,{\rm Hz}\left(\frac{\rho(R_*)}{10^{8}\,{\rm g\,cm^{-3}}}\right)^{1/2},
    \label{eq:fiNSBH}
\end{equation}
independent of the component masses.

Once formed near $R_*$, the
NS drifts inward by launching spiral density waves at its Lindblad resonances,
in analogy with Type~I planetary migration~\cite{Lin1986, Ward:1997di,Tanaka2002, 2010MNRAS.401.1950P, Kley2012}, on a
timescale~\cite{Baibhav:2026iny}
\begin{align}
    \tau_\text{\tiny mig}&\simeq 5\times10^{-2}\,{\rm s}
    \left(\frac{M_\BH}{5\,\msun}\right)
    \left(\frac{\mathbb{H}}{0.3}\right)^4 \nonumber \\
    & \times \left(\frac{m_\NS/M_\BH}{10^{-2}}\right)^{-1}
    \left(\frac{\dot{M}_\BH}{\msun\,{\rm s}^{-1}}\right)^{-1}\,.
    \label{eq:taumig}
\end{align}
Two dynamical regimes again emerge, now controlled mainly by the disk aspect
ratio and mass $M_\text{\tiny disk}$, and by whether the gas survives over the disk-depletion time
$\tau_\text{\tiny acc}^\BH \simeq10\,{\rm s}\,(M_\text{\tiny disk}/10\,\msun)(\dot{M}_\BH/\msun\,{\rm s}^{-1})^{-1}$,
until the NS reaches the inner disk. When
the central BH depletes the disk first, the NS completes its approach in vacuum,
driven purely by GW emission; when gas persists, the inspiral is
migration-assisted (Type~I) and, if hierarchical growth and accretion push the remnant
to heavier masses, it opens a gap and crosses into the slower Type~II
regime~\cite{Crida:2005zp, Kanagawa2018, Metzger:2024ujc,Baibhav:2026iny}. Any gas-driven torque or residual
eccentricity left at band entry is itself a signature distinguishing this
channel from a vacuum point-mass inspiral~\cite{Barausse:2014tra, Breivik:2016ddj, Nishizawa:2016eza, Cardoso:2019rou, Zevin:2021rtf, Romero-Shaw:2022xko, Zwick:2022dih, Wu:2026hth}, though for the background --- as in
the BNS case --- these enter only through the emission band and cycle count.
 
Regardless of the migration history, once the NS decouples from the gas, the
signal is that of a highly asymmetric compact binary whose chirp mass
$\mathcal{M}_c\simeq m_\NS^{3/5}M_\BH^{2/5}$ is enhanced by the more massive $(3-30)\,\msun$ BH companion. This makes the NS-BH merger the loudest source of the scenario, with a
horizon a factor $\sim3-10$ above the BNS one~\cite{Baibhav:2026iny}. Unbinding
plays no role here, so the inspiral terminates at the lower of two frequencies: the innermost stable circular orbit frequency of the (potentially spinning) central BH, and the Roche frequency at which the BH
tidally disrupts the NS~\cite{Vallisneri:1999nq, Pannarale:2011pk, Pannarale:2015jia}. The latter
is set by the EoS and depends only mildly on the BH mass,
\begin{equation}
    f_\text{\tiny Roche}^\text{\tiny NS-BH}
    =\frac{1}{\pi}\sqrt{\frac{m_\NS(1+q)}{8R_\NS^3}}
    \simeq 70\,{\rm Hz}\left(\frac{m_\NS}{0.1\,\msun}\right)\,.
    \label{eq:fRocheNSBH}
\end{equation}
Because the mass ratio suppresses the tidal
deformability below detectability~\cite{Baibhav:2026iny}, this disruption frequency is the channel's
primary spectral fingerprint, and it enters the background as a sharp,
mass-dependent cutoff. Equations~\eqref{eq:fi}--\eqref{eq:fRocheNSBH}, together
with the chirp mass, are the only per-source inputs required to build the
background in Sec.~\ref{sec:sgwb}.

\section{The stochastic background}
\label{sec:sgwb}
\noindent
The mergers of the two channels, summed over all redshifts and assumed isotropically distributed on the sky, source a SGWB whose energy density per logarithmic frequency, normalized to the
cosmological critical density $\rho_c=3H_0^2/8\pi$, reads~\cite{Phinney:2001di,Regimbau:2011rp}
\begin{widetext}
\begin{equation}
    \Omega_\GW(f)=\frac{f}{\rho_c}
    \int \! {\rm d}\bm{\theta}\;p(\bm{\theta})
    \int_0^{z_\text{\tiny max}(f,\bm\theta)}\!\!\!{\rm d}z\,
    \frac{\mathcal{R}(z)}{(1+z)\,H(z)}\,
    \frac{{\rm d}E_\GW}{{\rm d}f_s}\bigg|_{f_s=f(1+z)}\,,
    \label{eq:OmegaGW}
\end{equation}
\end{widetext}
where $H(z)= H_0[\Omega_r(1+z)^4+\Omega_m(1+z)^3+\Omega_\Lambda]^{1/2}$ is the Hubble rate, with $h=0.674$, $\Omega_m=0.315$,
$\Omega_\Lambda=0.685$, and $\Omega_r=5.38\times10^{-5}$~\cite{Planck:2018vyg}.
The factor $(1+z)^{-1}$ accounts for the redshifting of the emitted energy and
the time dilation of the rate, while $\mathcal{R}(z)$ is the source-frame merger
rate per comoving volume and $\bm{\theta}=(M_\BH,m_\NS)$ denotes the source masses. Each
source contributes at the redshifted frequency $f_s=f(1+z)$, so that a given
observed frequency $f$ collects emission from a range of redshifts bounded from above
by $z_\text{\tiny max}(f,\bm\theta)$, the redshift at which $f_s$ exceeds the source's
termination frequency and the emission leaves the band.
Equation~\eqref{eq:OmegaGW} gives the total background, including sources that may be individually resolved; for next-generation detectors, where many NS-BH events are resolvable (Table~\ref{tab:detect}), the residual after their subtraction would be lower, and the corresponding forecasts should be interpreted as upper bounds.
 
We model each merger as an inspiral, whose leading-order (Newtonian) energy
spectrum is
\begin{equation}
    \frac{{\rm d}E_\GW}{{\rm d}f_s}
    =\frac{\pi^{2/3}}{3}\,\mathcal{M}_c^{5/3}\,f_s^{-1/3}\,,
    \qquad f_i<f_s<f_\text{\tiny TD}\,,
    \label{eq:dEdf}
\end{equation}
with $\mathcal{M}_c=(m_1 m_2)^{3/5}/(m_1+m_2)^{1/5}$, band-limited between the
entry frequency $f_i$ [Eqs.~\eqref{eq:fi} and~\eqref{eq:fiNSBH}] and the termination frequency
$f_\text{\tiny TD}$ appropriate to each channel [Eqs.~\eqref{eq:fRocheBNS}
and~\eqref{eq:fRocheNSBH}]. Inserting Eq.~\eqref{eq:dEdf} into
Eq.~\eqref{eq:OmegaGW} yields the familiar $\Omega_\GW\propto f^{2/3}$ inspiral
scaling at frequencies well below $f_\text{\tiny TD}$, and it is precisely the band
limits that distinguish the collapsar background from that of ordinary compact
binaries. The anomalously low subsolar masses push $f_\text{\tiny TD}$ into the
most sensitive reach of ground-based detectors [Eqs.~\eqref{eq:fRocheBNS}
and~\eqref{eq:fRocheNSBH}], and the environmental terminator $f_\text{\tiny unb}$
(for the modulation case) can impose an additional hard cutoff with no counterpart
in a vacuum inspiral. 
We thus retain the leading-order vacuum energy spectrum, incorporating the source structure and environment only through the entry and termination frequencies. The resulting spectrum therefore rises as $f^{2/3}$, turns
over as the lightest sources reach termination, and cuts off sharply once the
heaviest do, establishing a shape controlled by the mass distribution and the EoS. All curves in Fig.~\ref{fig:sgwb} and all entries of Table~\ref{tab:detect} are obtained by evaluating the full population integral of Eq.~\eqref{eq:OmegaGW} separately for the NS-BH channel and for the two BNS regimes of Sec.~\ref{sec:bns}. The fragmentation density $\rho(R_*)$ is held fixed across the population, so that the remaining disk parameters ($\mathbb{H}$, $Q_0$, $\dot M_\BH$) are implicitly marginalized over, see discussion below. Since $\rho(R_*)$ sets the entry frequencies of Eqs.~\eqref{eq:fi} and~\eqref{eq:fiNSBH}, a larger value would shift the backgrounds to higher frequency, leaving the Roche cutoffs of Eqs.~\eqref{eq:fRocheBNS} and~\eqref{eq:fRocheNSBH} unchanged.
 
The volumetric rate inherits the collapsar formation history, which we tie to
the observed population of LGRBs. We assume~\cite{Baibhav:2026zuy} 
\begin{align}
    \mathcal{R}(z)&=f_m\,N_\NS\,\mathcal{R}_\text{\tiny LGRB}(z)\,, \nonumber \\
    \mathcal{R}_\text{\tiny LGRB}(z)&=\mathcal{R}_0\,
    \frac{(1+z)^{p_1}}{1+\big[(1+z)/p_2\big]^{p_3}}\,,
    \label{eq:Rate}
\end{align}
with $\mathcal{R}_0=79^{+57}_{-33}\,{\rm Gpc^{-3}\,yr^{-1}}$,
$p_1=3.33^{+0.33}_{-0.33}$, $p_2=3.42^{+0.13}_{-0.18}$, and
$p_3=6.21^{+0.38}_{-0.32}$, a parametrization that rises steeply at low redshift
and peaks near $z\simeq2$, tracking the cosmic history of massive-star
formation~\cite{Ghirlanda:2022edk}. We adopt the central values throughout, so
that the residual normalization uncertainty is absorbed into the free factor
$f_m$. The local normalization $\mathcal{R}_0$ is
the beaming-corrected LGRB rate~\cite{Ghirlanda:2022edk}, and the dimensionless
factor $f_m$ is the fraction of collapsars whose disks fragment and produce the
merger in question. It plays the same role as in the single-event rate estimate
of Ref.~\cite{Baibhav:2026zuy}, where $f_m\sim0.01-0.1$ was found to reproduce
the local rate of GW190814-like events; since $\mathcal{R}_\text{\tiny LGRB}$ counts only collapsars that launch successful jets, missing those whose jets are choked inside the star, it is a conservative lower bound on the true collapsar rate~\cite{Bromberg:2011wb}, and the factor $f_m$ takes into account both this and the fragmentation
efficiency. 

Each fragmenting collapsar seeds
$N_\NS\simeq2/\mathbb{H}\sim\mathcal{O}(10)$ disk-born NSs (Sec.~\ref{sec:disk}),
so the merger rate carries an additional multiplicity factor $N_\NS$; being
fully degenerate with $f_m$, only the product $f_m N_\NS$ is constrained by the
data, and throughout we quote bounds at the fiducial $N_\NS=10$, adopted as a representative value independently of the reference $\mathbb{H}=0.3$ used in the scaling relations of Sec.~\ref{sec:model}. Bounds for a different multiplicity follow by rescaling $f_m\to f_m\,(10/N_\NS)$. The value $N_\NS=10$ should be regarded as the maximal NS-BH multiplicity, attained when every fragment reaches the central BH individually. If some fragments first merged hierarchically with one another, fewer and heavier remnants would plunge into the BH. The NS-BH and BNS curves of Fig.~\ref{fig:sgwb}, evaluated at the same $f_m N_\NS$, should therefore be read as separate upper envelopes rather than as additive contributions.
The background bound is nonetheless robust to this bookkeeping: below the cutoff the energy radiated per NS-BH inspiral scales as $\mathcal{M}_c^{5/3}\simeq m_\NS M_\BH^{2/3}$, i.e., linearly in the NS mass, so that hierarchical growth conserves the total emitted energy while shifting the Roche cutoff of Eq.~\eqref{eq:fRocheNSBH}, for the polytropic EoS, to higher frequencies. 
Unlike ordinary compact binaries, whose merger rate follows the star-formation
history convolved with a broad delay-time distribution, the disk-born mergers
occur within seconds to years of the collapse itself [Eqs.~\eqref{eq:taudyn},
\eqref{eq:taumig}] --- negligible on cosmological timescales --- so that
$\mathcal{R}(z)$ tracks the LGRB rate essentially without time delay. This implies that the standard super-solar astrophysical populations, built from the star-formation
rate with a delay kernel~\cite{Virgo:2025aai}, peak at lower redshift than the
prompt collapsar population.
 
Finally, we take $p(\bm{\theta})$ to be uniform over $M_\BH\in[3,30]\,\msun$ and
$m_\NS\in[10^{-2},1]\,\msun$, reflecting our agnostic treatment of the highly
uncertain mass function of disk-born objects. 
We have verified that adopting instead a log-uniform prior on both masses, which gives more weight to the lightest fragments and BHs, leaves our results essentially unchanged.
Although Eq.~\eqref{eq:clumpmass} ties $m_\NS$ to $M_\BH$ at fixed disk parameters, the aspect ratio $\mathbb{H}$, the Toomre threshold $Q_0$, and the accretion rate $\dot M_\BH$ vary from one collapsar to another and are themselves poorly constrained. The independent prior on $m_\NS$ should therefore be understood as implicitly marginalizing over them. For instance, $\mathbb{H}\simeq 0.1-0.5$ with $Q_0\simeq\mathcal{O}(1)$ maps the BH mass window onto $m_\NS\simeq \mathcal{O}(10^{-3}-1)\,\msun$ through Eq.~\eqref{eq:clumpmass}, comfortably covering the adopted range. 
Because
$\mathcal{R}(z)\propto f_m N_\NS$ and because it enters Eq.~\eqref{eq:OmegaGW} linearly, the
background amplitude scales simply as $\Omega_\GW\propto f_m N_\NS$, so that the
confrontation with observations in Sec.~\ref{sec:results} translates directly
into a constraint on the collapsar fraction.
 
\subsection*{Statistical regime and detectability}
\noindent
Whether the background is a continuous Gaussian signal or an intermittent
``popcorn'' one is controlled by the duty cycle, the mean number of
events overlapping in band~\cite{Regimbau:2008nj,Rosado:2011kv,Lawrence:2023buo},
\begin{equation}
    \xi=\int\!{\rm d}\bm\theta\,p(\bm\theta)\!\int_0^{z_\text{\tiny max}}\!\!\!{\rm d}z\,
    \mathcal{R}(z)\,\bar\tau_s(\bm\theta)\,\frac{{\rm d}V_c}{{\rm d}z}\,,
    \quad
    \frac{{\rm d}V_c}{{\rm d}z}=\frac{4\pi\,r^2(z)}{H(z)}\,,
    \label{eq:duty}
\end{equation}
where $r(z)=\int_0^z{\rm d}z'/H(z')$ is the comoving distance, and $z_\text{\tiny max}=f_\text{\tiny TD}/f_\text{\tiny det,min}-1$ is the band-exit redshift of Eq.~\eqref{eq:OmegaGW}. The source-frame time spent in band is $\bar\tau_s=\tau_\GW(f_\text{\tiny low})-\tau_\GW(f_\text{\tiny TD})$, with $\tau_\GW(f)=\tfrac{5}{256}(\pi f)^{-8/3}\mathcal{M}_c^{-5/3}$ the Newtonian time to coalescence. Its lower edge, $f_\text{\tiny low}=\max[f_i,\,f_\text{\tiny det,min}(1+z)]$, is set either by the source entry frequency or by the detector low-frequency limit $f_\text{\tiny det,min}$, redshifted to the source frame. The magnitude of $\xi$ fixes the statistical character of the signal: for
$\xi\gg1$ many sources overlap in band at once and the background is a continuous,
Gaussian signal; for $\xi\ll1$ it degrades into an intermittent ``popcorn'' train
of well-separated events. The
source-frame time in band scales as $\tau_\GW\propto\mathcal{M}_c^{-5/3}$, so
lighter binaries last longer and, at fixed rate and reachable volume, push
$\xi$ upward. This ordering is familiar from the standard astrophysical
backgrounds: the stellar-mass binary BH population is deeply intermittent
($\xi\sim10^{-3}$~\cite{Lawrence:2023buo}), whereas the lighter binary NS
population sits closer to the continuous regime. The duty cycles of the two
collapsar channels, with the BNS one bracketed by its two regimes, quoted in Sec.~\ref{sec:results}, follow from the same
interplay of in-band duration, merger rate, and reachable volume.

\begin{figure*}[t!]
    \centering
    \includegraphics[width=0.8\linewidth]{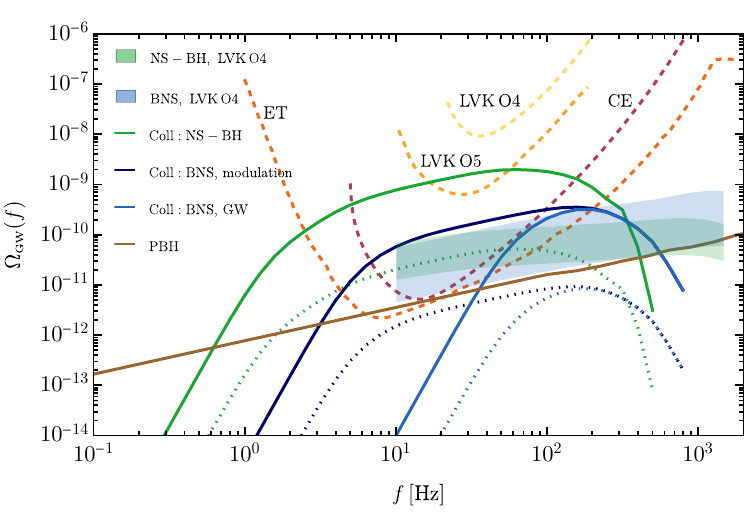}
    \caption{Stochastic GW background from subsolar mergers in collapsar disks, for the polytropic EoS at $\rho(R_*)=10^{8}\,{\rm g\,cm^{-3}}$ and $f_\text{\tiny gas} = 10^{-4}$. Green curves show the NS-BH channel, dark and light blue the BNS modulation and GW regimes; for each curve the solid line corresponds to the bound from the non-detection of the NS-BH background ($S/N<3$) and the dotted line to the tighter bound attainable by a dedicated search for resolvable NS-BH events ($N_\text{\tiny det}<1$) in O4. The brown line shows the SGWB from PBH binaries for a narrow lognormal mass function centered at $0.1\,\msun$ with abundance $f_\PBH=10^{-3}$, at the edge of current microlensing and subsolar GW limits. The dashed curves are power-law-integrated sensitivities: for LVK, the $2\sigma$ curves of Ref.~\cite{Virgo:2025aai} from the O1--O4a data (labeled O4) and for one year at the O5 target sensitivity; for ET and the two-site CE network, the curves of Ref.~\cite{Branchesi:2023mws} for $T=1 \, {\rm yr}$ and $S/N=1$. The shaded bands are the standard astrophysical NS-BH (green) and BNS (blue) backgrounds forecast for O4~\cite{Virgo:2025aai}, computed for the ordinary compact-binary mass range, that overlap the collapsar signal in amplitude but are distinguished from it by the characteristic turnover
and cutoff imposed by the low subsolar masses.}
    \label{fig:sgwb}
\end{figure*}

In the continuous regime the optimal search is the standard cross-correlation estimator. Treating each network as a single L-shaped, colocated and coaligned detector, i.e., neglecting the overlap reduction function $\gamma(f)$ of the actual detector pairs, the accumulated signal-to-noise ratio (SNR, $S/N$) over an observation time $T$ takes the compact form~\cite{Allen:1997ad,Thrane:2013oya}
\begin{equation}
    S/N=\frac{\sqrt{2T}}{5}\left[\int_{f_\text{\tiny det,min}}^{f_\text{\tiny det,max}}\!{\rm d}f\,
    \left(\frac{\Omega_\GW(f)}{\Omega_\text{\tiny det}(f)}\right)^{\!2}\right]^{1/2}\,,
    \label{eq:snr}
\end{equation}
where the integral runs over the sensitive band $[f_\text{\tiny det,min},f_\text{\tiny det,max}]$ of the instrument, and $\Omega_\text{\tiny det}(f)=(2\pi^2/3H_0^2)\,f^3 S_n(f)$ is the noise energy-density spectrum built from the detector strain power spectral density $S_n(f)$. This idealized limit, adopted for all detectors considered here, overestimates the cross-correlation $S/N$ by a factor of order $1/|\gamma|$. 
The sensitivity curves shown in Fig.~\ref{fig:sgwb} are instead the power-law-integrated curves~\cite{Thrane:2013oya} computed for the actual detector networks, including their overlap reduction function~\cite{Virgo:2025aai,Branchesi:2023mws}. We adopt minimum frequencies $f_\text{\tiny det,min}=20$, $2$, and $5$~Hz for LVK O4, ET, and CE, respectively. The downward shift of the low-frequency edge from O4 to the next-generation instruments is largely responsible for the growth of the SNR.

\paragraph*{Resolvable events.}
The same population also produces individual coalescences that a given
instrument may resolve, an observable complementary to the stochastic background
above. Defining the optimal SNR
$\rho_\text{\tiny opt}(\bm\theta,z)$ as the value attained for a binary directly
overhead and face-on, the actual source SNR is $\rho=w\,\rho_\text{\tiny opt}$, where
$w\in[0,1]$ is the projection factor fixed by the sky position, polarization,
and inclination through the antenna pattern functions. The source is
detected when $\rho>\rho_\text{\tiny thr}$, so that its detection probability is
the survival function of $w$,
\begin{equation}
    p_\text{\tiny det}(\bm\theta,z)
    =P\!\left(w>\frac{\rho_\text{\tiny thr}}{\rho_\text{\tiny opt}(\bm\theta,z)}\right),
    \label{eq:pdet}
\end{equation}
which we evaluate with a threshold
$\rho_\text{\tiny thr}=8$, following Appendix~A of Ref.~\cite{DeLuca:2021wjr}. The
number of resolvable events collected over an observation time $T$ is then~\cite{DeLuca:2021hde}
\begin{equation}
    N_\text{\tiny det}=T\int\!{\rm d}\bm\theta\,p(\bm\theta)
    \int_0^{z_\text{\tiny max}}\!\!\!{\rm d}z\,
    \frac{\mathcal{R}(z)}{1+z}\,\frac{{\rm d}V_c}{{\rm d}z}\,
    p_\text{\tiny det}(\bm\theta,z)\,.
    \label{eq:Ndet}
\end{equation}
Equations~\eqref{eq:snr}
and~\eqref{eq:Ndet} give two complementary criteria to confront the collapsar
prediction with a null result in a given run. The non-detection of an isotropic
background requires the predicted $S/N<3$, a bound that rests only on the
integrated energy density; the absence of any resolved subsolar coalescence
requires $N_\text{\tiny det}<1$, a tighter but more model-dependent bound that
relies on the single-event detection modeling. Since both $\Omega_\GW$ and
$N_\text{\tiny det}$ scale linearly with $f_m N_\NS$, each criterion maps
directly into an upper bound on the collapsar fraction. Representative values of
$\xi$, $S/N$, and $N_\text{\tiny det}$, together with the resulting bounds, are
quoted in Sec.~\ref{sec:results}.

\begin{table*}[t!]
\centering
\renewcommand{\arraystretch}{1.3}
\begin{tabular}{l c c c c c c}
\hline\hline
 & \multicolumn{3}{c}{$S/N$} & \multicolumn{3}{c}{$N_\mathrm{det}$} \\
\cline{2-4}\cline{5-7}
Channel & LVK O4 & ET & CE & LVK O4 & ET & CE \\
\hline
NS-BH            & $7 \times 10^{-1}$             & $5\times10^{2}$ & $8\times10^{2}$ & $9$              & $8\times10^{4}$ & $2\times10^{5}$ \\
BNS, GW regime         & $3\times10^{-2}$   & $3$ & $6$  & $2\times10^{-2}$  & $4\times10^{1}$            & $6 \times 10^1$ \\
BNS, modulation regime & $8 \times 10^{-2}$   & $3\times10^{1}$  & $7\times10^{1}$   & $3\times10^{-1}$             & $1\times10^{3}$           & $7\times10^{3}$ \\
\hline\hline
\end{tabular}
\caption{Background signal-to-noise ratio $S/N$ and number of resolvable events $N_\text{\tiny det}$ for the two collapsar channels, with the BNS one shown in both the GW and modulation regimes, in LVK O4, ET, and CE, for an observation time $T=1\,{\rm yr}$ and reference normalization $f_m N_\NS=1$. Both scale linearly with $f_m N_\NS$; note in addition that $N_\text{\tiny det}\propto T$ and $S/N\propto\sqrt{T}$. Entries are rounded to one significant figure, while the bounds quoted in the text are computed from the unrounded values. 
}
\label{tab:detect}
\end{table*}

\section{Results and current bounds}
\label{sec:results}
\noindent
Figure~\ref{fig:sgwb} shows the background sourced by the two channels, with the BNS one shown in both regimes,
computed from Eqs.~\eqref{eq:OmegaGW}--\eqref{eq:Rate} for the polytropic EoS at
the fiducial disk density $\rho(R_*)=10^{8}\,{\rm g\,cm^{-3}}$ and $f_\text{\tiny gas} = 10^{-4}$. For each curve,
the solid and dotted line styles correspond to the two O4 bounds on
$f_m N_\NS$ derived below. All spectra share the same qualitative shape dictated
by Eq.~\eqref{eq:dEdf}: an inspiral rise $\Omega_\GW\propto f^{2/3}$ at low
frequency, a turnover as the lightest sources begin to reach their termination
frequency, and a sharp high-frequency cutoff once even the heaviest do. The
location of the turnover and cutoff is therefore a direct imprint of the mass
distribution and of the EoS.
Fig.~\ref{fig:sgwb} also shows the standard astrophysical NS-BH and BNS backgrounds forecast for O4~\cite{Virgo:2025aai}, computed for the ordinary compact-binary mass range: they overlap the collapsar signal in amplitude, but lack its low-mass turnover and cutoff.
Below the turnover the spectra are well approximated by $\Omega_\GW(f)\simeq A_\GW\,f_mN_\NS\,(f/25\,{\rm Hz})^{2/3}$, with $A_\GW\simeq 3.2 \times 10^{-10}$ for the NS-BH channel and $A_\GW\simeq 3 \times 10^{-13}$ and $3 \times 10^{-11}$ for the BNS channel in the modulation and GW regimes, respectively.

The hierarchy among the channels is mostly determined by the source-level chirp-mass. The NS-BH background is the loudest, peaking around
a few tens of Hz, while the BNS channel lies about one to two orders of magnitude
below, the gap inherited from the source level where the BH-dominated chirp mass
makes ${\rm d}E_\GW/{\rm d}f_s\propto\mathcal{M}_c^{5/3}$ larger than for the symmetric
BNS. Between the BNS regimes, the modulation regime peaks at lower
frequency than the GW one, for two reasons: the modulation regime requires wide
inner binaries (large $f_\text{\tiny H}$), which through
$f^\BNS_i\propto f_\text{\tiny H}^{-3/2}$ [Eq.~\eqref{eq:fi}] shifts the entry
frequency downward, and the potential early unbinding of the pair
($f_\text{\tiny unb}<f_\text{\tiny Roche}$) could truncate the emission before
contact, whereas the GW regime enters higher and radiates all the way to the contact frequency. These peak locations depend on the residual gas fraction, fixed here to $f_\text{\tiny gas}=10^{-4}$. A larger $f_\text{\tiny gas}$ shortens $\tau_\text{\tiny dyn}$ [Eq.~\eqref{eq:taudyn}] and enlarges the gas-driven region, so that the GW-driven regimes are restricted to tighter inner binaries (smaller $f_\text{\tiny H}$), which enter the band at higher frequency. This shifts the BNS backgrounds to higher frequencies with comparable amplitude.
At the background $S/N<3$ bound the BNS channel lies one to two orders of magnitude
below O4, but falls within reach of next-generation detectors; the NS-BH channel
remains the loudest and is accessible to both current and future experiments.

The detectability of each channel is quantified in Table~\ref{tab:detect}, which collects the background SNR and the number of resolvable events in LVK O4, ET, and CE, for $T=1\,{\rm yr}$ and the reference normalization $f_m N_\NS=1$; both scale linearly with $f_m N_\NS$. The duty cycle of the loud NS-BH channel scales as $\xi\propto f_m N_\NS$ like
the amplitude; at the O4 background bound (see below)
it is $\xi_{\NS-\BH}\simeq \mathcal{O}(1)$ in O4 and $\xi_{\NS-\BH} \simeq \mathcal{O}(6 \times 10^2)$ for ET, so the background
is firmly continuous in ET and of order unity --- marginally continuous --- at
current sensitivity, while the BNS channel splits: the GW regime, populated only by tight binaries entering at hundreds of Hz, has $\xi_\BNS\lesssim1$, whereas the modulation regime, whose wide binaries remain in band for hours, reaches $\xi_\BNS\simeq\mathcal{O}(30)$ in O4 and $\mathcal{O}(10^2)$ in ET.  Since $\xi\propto f_mN_\NS$, near the projected sensitivity of next-generation detectors the duty cycle drops correspondingly, and a search statistic accounting for non-Gaussianity may be preferable (see e.g.~\cite{Zhong:2026jzl}).

Consider first the background $S/N$, whose hierarchy mirrors that of the amplitude, with the NS-BH channel one to two orders of magnitude above the BNS one. From O4 to the next-generation instruments the $S/N$ of every curve grows by two to three orders of magnitude, driven by the sharply lower noise floor, with CE improving on ET by a further factor of a few in each case. Unlike $N_\text{\tiny det}$, which is dominated by the loudest individual sources at the low-frequency end of the band, the background
$S/N$ accumulates where $\Omega_\GW$ overlaps best with the sensitivity curve, in
the decade around a few tens of Hz near the spectral peak fixed by
the subsolar termination frequencies. The modulation regime, peaking closer to the minimum of $\Omega_\text{\tiny det}$, reaches an SNR few times larger than the GW regime at O4, widening to a factor of $\sim 10$ for ET and CE. It is the NS-BH value that fixes
the background bound $f_m\lesssim0.45$ of Eq.~\eqref{eq:fmbound} below.

The two BNS regimes, however, part ways in their number of resolvable events. Although both draw on the same mass and redshift distribution, the single-source SNR accumulates mostly at the low-frequency end of the inspiral, so the band edges that distinguish them do feed into $p_\text{\tiny det}$. Its lower entry frequency lets the modulation regime capture more of this portion, giving an $N_\text{\tiny det}$ about one to two orders of magnitude larger than the GW regime at all three detectors. In absolute terms both regimes are unresolvable at O4, where the small subsolar chirp mass keeps each source faint, but the situation reverses at ET and CE: their lower $f_\text{\tiny det,min}$ and strain floor recover the low-frequency portion of the inspiral that these light sources emit most loudly. The BNS channel thus
turns from individually invisible at O4 into a rich resolvable population at ET,
while staying subdominant to NS-BH in the background amplitude.
 
The most immediate consequence follows from confronting the NS-BH prediction with
the non-detection in the first part of the LVK O4 run~\cite{Virgo:2025aai}.
First, the absence of a detectable isotropic background requires the predicted
SNR to satisfy $S/N<3$; since $S/N\propto f_m N_\NS$, this caps
the amplitude and yields a bound resting only on the integrated energy
density. Second, the non-detection of any resolved subsolar coalescence
requires $N_\text{\tiny det}<1$. This bound is tighter but conditional:
it assumes a genuine null result, whereas a few of the marginal candidates
reported so far could, in principle, be compatible with the collapsar
scenario~\cite{Christensen:2026qfx}. Evaluating both criteria for the
NS-BH channel gives
\begin{align}
    f_m &\lesssim 0.45 & &(S/N<3)\,, \nonumber \\
    f_m &\lesssim 1.2\times10^{-2} & &(N_\text{\tiny det}<1)\,,
    \label{eq:fmbound}
\end{align}
at the fiducial $N_\NS=10$, corresponding respectively to the solid and dotted curves in Fig.~\ref{fig:sgwb}. For O4, a further comparison uses the 95\% upper limit on an $f^{2/3}$ background~\cite{Virgo:2025aai}: requiring $\Omega_\GW(25\,{\rm Hz})<2\times10^{-9}$ (log-uniform prior) gives $f_m\lesssim 0.63$, compatible with the estimate above. The event-based bound, instead, assumes an optimal matched-filtering search covering the full collapsar parameter space. Existing subsolar template banks span only a restricted range of component masses and mass ratios~\cite{LIGOScientific:2021job,LVK:2022ydq,LIGOScientific:2026wxz}, and do not include some of the extreme mass ratios typical of the NS-BH channel. This bound should therefore be read as the sensitivity that a dedicated search on O4 data would achieve, rather than as a constraint implied by current null results, and motivates extending subsolar searches to this region of parameter space~\cite{Cheung:2025grp}. The background bound, which requires no template coverage, is instead a genuine constraint from current data.

The two bounds are mutually consistent: at the
background bound $f_m\simeq0.45$ one already expects tens of resolvable NS-BH events, which is precisely why the event criterion pushes the
bound an order of magnitude lower. The tighter, event-based bound is competitive
with the range $f_m\sim0.01-0.1$ invoked in Ref.~\cite{Baibhav:2026zuy} to
reproduce the local rate of GW190814-like events. Conversely, if the subsolar candidates S250818k and S251112cm were of astrophysical origin and belonged to the NS-BH channel, one or two events in O4 would correspond to $f_m\simeq \mathcal{O}(10^{-2})$, at the lower end of that range.

Both bounds scale linearly with the local rate: across the $1\sigma$ range of $\mathcal{R}_0$ they span $f_m\lesssim0.27-0.78$ and $f_m\lesssim(0.7-2)\times10^{-2}$, respectively, while the uncertainties on $p_{1,2,3}$, which only reshape the redshift distribution, have a subleading impact.
Because $f_m$ absorbs the fragmentation efficiency and
$\mathcal{R}_\text{\tiny LGRB}$ is itself a lower bound on the collapsar rate, the
true constraint on the fraction of \emph{fragmenting} collapsars may be tighter
still. These bounds are also insensitive to the residual gas fraction, since the NS-BH channel that sets them does not depend on $f_\text{\tiny gas}$.

A natural point of comparison is the SGWB expected from PBHs, the standard
subsolar candidate introduced above. The brown curve in Fig.~\ref{fig:sgwb}
shows the background from PBH binaries, taken from a narrow lognormal mass function
centered at $0.1\,\msun$ and cosmological abundance with respect to the dark matter $f_\PBH=10^{-3}$, fixed at the edge of
current microlensing and subsolar GW bounds~\cite{Carr:2020gox,Andres-Carcasona:2024wqk}, with
the merger rate of the standard early-Universe pairing
channel~\cite{Sasaki:2016jop,Raidal:2018bbj,Raidal:2024bmm}. Being
a population of point masses in vacuum, it radiates a pure $f^{2/3}$ inspiral
spectrum with no EoS-driven turnover or termination cutoff, rising monotonically
across the band, and reaches $S/N\simeq \mathcal{O}(10^{-2})$ at O4 and $\mathcal{O}(10)$ at ET. At the
allowed abundance its amplitude is partly below 
the collapsar NS-BH signal, so the two may be distinguished both by amplitude and shape, since the collapsar background turns over and cuts off sharply in the most sensitive
decade, whereas the PBH background continues as an unbroken power law. The two
have comparable duty cycles, so this spectral difference, rather than the
statistics of the background, provides the cleaner discriminant between primordial
and disk-born subsolar objects.

Looking ahead, the projected power-law-integrated sensitivities of ET
and CE~\cite{Branchesi:2023mws,Evans:2021gyd} lie well below the NS-BH prediction over part of the band, as quantified by the integrated SNR of Table~\ref{tab:detect}. The next generation experiments will therefore either detect this background or tighten the bound of Eq.~\eqref{eq:fmbound} substantially: for ET the background criterion alone would reach  $f_m\lesssim6\times10^{-4}$ and the resolvable-event one $f_m\lesssim10^{-6}$, tightening at CE to $f_m\lesssim4\times10^{-4}$ and $f_m\lesssim6\times10^{-7}$, an improvement of several orders of magnitude over O4. A detection would be corroborated by the distinctive
spectral shape, which would encode the characteristic subsolar mass scale of the
fragments directly in the spectrum. We have adopted the polytropic EoS throughout;
varying it leaves the low-frequency $f^{2/3}$ tail essentially unchanged, as
expected, while the high-frequency cutoff shifts through $f_\text{\tiny Roche}\propto R_\NS^{-3/2}$, as discussed for the EoSs of Ref.~\cite{Baibhav:2026iny}. 
These background forecasts are idealized: at $f_m\sim10^{-3}$ the collapsar signal lies below the astrophysical compact-binary background that ET and CE will detect at high significance, so that the $S/N<3$ criterion is attainable only after subtraction of resolved sources and a joint spectral fit separating the collapsar component from the residual foreground~\cite{Cutler:2005qq, Sharma:2020btq, Sachdev:2020bkk, Biscoveanu:2020gds, Zhou:2022nmt, Zhong:2024dss, Zhong:2025qno, Clarke:2026zed}. The same caveat applies to the spectral discrimination discussed above, which must survive the uncertainties in the high-frequency shape of the foreground itself. The resolvable-event bounds, relying on individual detections, are not affected.

In addition to mergers, the self-gravitating disk formed during the collapse of the massive star is prone to non-axisymmetric deformations, whose time-varying quadrupole moment sources a GW burst~\cite{Lam:2026asp} with characteristic frequency and strain
\begin{align}
    f_\text{\tiny axi}& = \frac{\Omega(R_\text{\tiny disk})}{\pi}
    \simeq 2\,{\rm Hz}\,
    \left(\frac{R_\text{\tiny disk}}{10^2 R_g}\right)^{-3/2}
    \left(\frac{M_\BH}{30\msun}\right)^{-1}\,,
    \nonumber \\
    h_\text{\tiny axi}
    & \simeq \frac{4\,\varepsilon\,M_\text{\tiny disk}\,R_\text{\tiny disk}^2\,\Omega^2(R_\text{\tiny disk})}{d_\text{\tiny L}}
    \nonumber \\
    & \simeq 2\times10^{-25}
    \left(\frac{\varepsilon}{0.1}\right)
    \left(\frac{M_\text{\tiny disk}}{\msun}\right)
    \left(\frac{R_\text{\tiny disk}}{10^2 R_g}\right)^{-1}
    \left(\frac{d_\text{\tiny L}}{\rm Gpc}\right)^{-1}\,,
    \label{eq:collapse}
\end{align}
where $\Omega(R_\text{\tiny disk})$ is the orbital frequency at the disk outer radius $R_\text{\tiny disk}$, $\varepsilon\leq1$ is the fraction of the disk mass $M_\text{\tiny disk}$ carried by the non-axisymmetric deformation, and $d_\text{\tiny L}$ the luminosity distance. The emission sits in the deci-Hz band accessible to DECIGO~\cite{Kawamura:2020pcg}, and --- summed over the collapsar population --- would source a stochastic background of its own. Its amplitude is throttled by the short disk lifetime. Because GW backreaction is weak, with a spin-down time $\tau_\GW^\text{\tiny axi}=f_\text{\tiny axi}/\dot f_\text{\tiny axi}\sim\mathcal{O}(\text{day})$ far longer than the depletion time $\tau_\text{\tiny acc}^\BH\sim\mathcal{O}(10)\,$s over which the central BH consumes the disk, each source radiates for only $\tau_\text{\tiny acc}^\BH$ at the quadrupole luminosity~\cite{Maggiore:2007ulw}
\begin{equation}
    \dot E_\GW^\text{\tiny axi}\simeq\frac{32}{5}\,\varepsilon^2\,M_\text{\tiny disk}^2\,R_\text{\tiny disk}^4\,\Omega^6(R_\text{\tiny disk})\,,
    \label{eq:dEdf_axi}
\end{equation}
so that the energy spectrum is approximately ${\rm d}E_\GW/{\rm d}f_s \simeq \dot E_\GW^\text{\tiny axi} \,\tau_\text{\tiny acc}^\BH\, \delta\big(f_s -f_\text{\tiny axi}\big)$. Inserting this into Eq.~\eqref{eq:OmegaGW} yields the amplitude
\begin{equation}
\Omega^\text{\tiny axi}_\GW (f) \simeq \frac{\dot E_\GW^\text{\tiny axi} \,\tau_\text{\tiny acc}^\BH}{\rho_c} \, \frac{f_m\,\mathcal{R}_\text{\tiny LGRB}(z)}{(1+z)\,H(z)} \bigg|_{z = f_\text{\tiny axi}/f - 1}\,,
\end{equation}
which, at the characteristic frequency of $f_\text{\tiny axi} \simeq 1 \, {\rm Hz}$, evaluates to
\begin{equation}
\Omega^\text{\tiny axi}_\GW(1 \, {\rm Hz}) \simeq 3 \times 10^{-15} \left(\frac{\varepsilon}{0.1}\right)^2 \left(\frac{f_m}{10^{-1}}\right)\,,
\end{equation}
where, since each collapsar hosts a single disk, the rate carries no multiplicity factor $N_\NS$. Although every collapsar forms a disk, the relevant prefactor is the fraction of disks developing a sizable non-axisymmetric deformation; since the disks that fragment are precisely the self-gravitating ones prone to such instabilities, this fraction is at least $f_m$, and adopting $f_m$ yields a conservative estimate. We have also set $M_\text{\tiny disk}\simeq0.8\,M_\BH$ and $R_\text{\tiny disk}\simeq R_*$, integrating over the BH mass distribution of Sec.~\ref{sec:sgwb}. This signal is well below the merger channels; we therefore quote it as an analytic estimate rather than showing it as a curve in Fig.~\ref{fig:sgwb}. Nevertheless, DECIGO reaches $S/N \simeq 5$ at the background bound $f_m = 0.45$, with a duty cycle of $\xi \simeq 1$. This signal thus offers an independent GW probe of the same collapsar systems, with its loudness set by the interplay between the disk parameters and the accretion timescale.

\section{Conclusions}
\label{sec:conclusions}
\noindent
We have computed the SGWB produced by the subsolar NSs that
form in fragmenting collapsar disks, treating both merger channels of the
scenario: the mergers of two disk-born NSs and the inspiral of a surviving NS
remnant into the central BH. Tying the formation history to the observed LGRB
rate and folding in the entry and termination frequencies set by the NS
EoS and by the disruptive tidal field of the central BH, we find that the NS-BH
channel is the loudest, peaking at most near $\Omega_\GW\sim10^{-9}$, while the BNS
channel lies up to about two orders of magnitude below. The NS-BH channel is already constrained at current sensitivity and, together with the BNS channel in the modulation regime, falls within reach of next-generation observatories across the allowed range, while the BNS GW regime becomes accessible only near the background bound.

Confronting the prediction with the O4 non-detection provides two complementary,
mutually consistent bounds on the fraction of collapsars that source these
mergers: one from the absence of a detectable background, $f_m\lesssim0.5$,
and a tighter one attainable by a dedicated search for resolvable subsolar events,
$f_m\lesssim10^{-2}$ (at the fiducial multiplicity $N_\NS=10$). Indeed, at the background bound one would already expect tens of resolvable NS-BH events in O4, which is why the single event criterion is an order of magnitude tighter. The latter, however, assumes a genuine null result in a dedicated search for subsolar candidates, whereas the former rests on a distinct observable and holds independently of individual-event searches. A dedicated search would probe the range invoked to reproduce the local rate of GW190814-like events~\cite{Baibhav:2026zuy}. Both will improve by three to four orders of magnitude with next-generation detectors, with the background search depending on our ability to separate foregrounds~\cite{Cutler:2005qq, Sharma:2020btq, Sachdev:2020bkk, Biscoveanu:2020gds, Zhou:2022nmt, Zhong:2024dss, Zhong:2025qno, Clarke:2026zed}. The background carries a distinctive spectral fingerprint
--- an inspiral $f^{2/3}$ rise terminated by a sharp, EoS-dependent cutoff driven
into the most sensitive band by the anomalously low masses --- that separates it
from the smooth compact-binary background and offers a population-level probe of
the collapsar channel, complementary to the individual-event signatures of the
companion analysis~\cite{Baibhav:2026iny}.

Several directions are worth pursuing. The result depends on the poorly known mass function of
disk-born objects, which we have taken to be uniform, although a log-uniform prior yields similar results; a physically motivated
distribution, informed by fragmentation simulations~\cite{Chen:2025uwd, Wu:2026hth}, would sharpen both the
amplitude and the bound on $f_m$.  
On the observational side, turning the event-based sensitivity into an actual constraint requires dedicated searches tailored to the collapsar channel: subsolar template banks should be extended to the extreme mass ratios $q\sim10^{-3}-10^{-1}$ of the NS-BH systems formed in these disks~\cite{Cheung:2025grp}, and should include the residual eccentricity that gas-driven migration and hierarchical-triple dynamics may leave when binaries enter the observable band~\cite{Baibhav:2026iny} (see also Refs.~\cite{Lynch:2026ibo,Islam:2026eev,Gamboa:2026ybr}). Such searches would also directly test whether marginal subsolar candidates are compatible with a disk-born origin.
Because the sources trace the LGRB population,
the background is expected to be anisotropic and partially correlated with the
large-scale distribution of collapsars, making it a target for directional and
cross-correlation searches with electromagnetic catalogs. Beyond the mergers, we have estimated the distinct low-frequency background sourced by the non-axisymmetric deformation of the nascent disk: peaking in the deci-Hz band within reach of DECIGO, it offers an independent probe of the same systems, whose detailed characterization --- with a realistic treatment of the mode amplitudes and emission duration --- we leave to future work. Finally, the long-lived,
quasi-monochromatic $f$-mode oscillation of the subsolar merger remnant, shifted
to $\sim100\,$Hz~\cite{Andersson:1997rn, Vretinaris:2019spn, Lioutas:2017xtn, Baibhav:2026iny}, could add a narrowband feature to the
background beyond the inspiral contribution computed here. A joint interpretation
of resolved subsolar events and of this stochastic background would tie together
accretion physics, the neutron-star EoS, and the collapsar formation rate within
a single observational program.

\section*{Acknowledgments}
\noindent
We acknowledge interesting discussions with Vishal Baibhav, Nelson Christensen, Brian D. Metzger, Lam Hui, Francesco Iacovelli, Luca Reali and Haowen Zhong. E.B, V.D.L. and L.D.G. are supported by NSF Grants No.~AST-2606672, No.~PHY-2513337, No.~PHY-090003, and No.~PHY-20043, by the Simons Foundation [MPS-SIP-00001698, E.B.], by the Simons Foundation International [SFI-MPS-BH-00012593-02], and by Italian Ministry of Foreign Affairs and International Cooperation Grant No.~PGR01167.
This work was carried out at the Advanced Research Computing at Hopkins (ARCH) core facility (\url{https://www.arch.jhu.edu/}), which is supported by the NSF Grant No.~OAC-1920103.

\bibliography{draft}

\end{document}